\documentclass[letterpaper,10pt]{article}

\usepackage{sectsty}
\usepackage[T1]{fontenc}
\usepackage{mathptmx}
\usepackage[round]{natbib}
\usepackage{graphicx}
\usepackage[font=footnotesize,labelfont={bf,sf},textfont=sf,singlelinecheck=0,justification=raggedright]{caption}
\title{Transmission spectroscopy of the sodium `D' doublet in {WASP-17b} with the {VLT}\footnote{Based on observations made with ESO Telescopes at the Paranal Observatory under programme ID 083.C-0292.}}

\author{ Patricia L Wood\footnote{Email: plw@astro.keele.ac.uk}, Pierre F L Maxted, Barry Smalley \& Nicolas Iro\\
\small Astrophysics Dept., Keele University, Staffordshire ST5 5BG, UK}

\date{}

\begin{document}

\allsectionsfont{\small\bfseries}

\maketitle

\begin{abstract}
\normalsize The detection of increased sodium absorption during primary transit implies the presence of an atmosphere around an extrasolar planet, and enables us to infer the structure of this atmosphere. {\sloppy Sodium has only been detected in the atmospheres of two planets to date -- HD\,189733b and HD\,209458b.  WASP-17b is the least dense planet currently known. It has a radius approximately twice that of Jupiter and orbits an F6-type star.} The transit signal is expected to be about five times larger than that observed in HD\,209458b. We obtained 24 spectra with the GIRAFFE spectrograph on the VLT, eight during transit. The integrated flux in the sodium doublet at wavelengths 5889.95 and 5895.92\,{\AA} was measured at bandwidths 0.75, 1.5, 3.0, 4.0, 5.0, and 6.0\,{\AA}. We find a transit depth of $0.55\pm0.13$ per cent at 1.5\,{\AA}. This suggests that, like HD\,209458b, {WASP-17b} has an atmosphere depleted in sodium compared to models for a cloud-free atmosphere with solar sodium abundance. We observe a sharp cut-off in sodium absorption between 3.0 and 4.0\,{\AA}, which may indicate a layer of clouds high in the atmosphere.
\end{abstract}

%\begin{multicols}{2}

\small
\section{Introduction}\label{sec:in}
\par Transiting extrasolar planets present an opportunity to study the properties of planetary atmospheres using transmission spectroscopy. The atmosphere regulates heat transfer around the planet, and determines what fraction of the incoming flux from the star is absorbed \citep{Sho02}. This can have a dramatic effect on the structure and evolution of a hot Jupiter; e.g. additional heating can cause the planet to expand, increasing the strength of the tidal interaction with its star. This may result in the planet's destruction as it spirals into its host star \citep*{Lev09}.

\par When light from a star passes through the atmosphere of a transiting planet, some wavelengths are absorbed much more strongly than others, so the apparent radius of the planet is larger at these wavelengths. The base of the atmosphere of a Jupiter-like planet is generally defined as the atmospheric depth at which the slant optical depth is equal to 2/3 \citep{Bur09}. At the typical temperature of a hot Jupiter, strong absorption is predicted in the sodium doublet line at wavelengths 5889.95 and 5895.92\,{\AA} \citep{Sea00}. The ratio of the area of a hot Jupiter's atmosphere to the disc of its star is typically $10^{-4}-10^{-3}$, so during transit the depth of the sodium absorption lines relative to the continuum are expected to increase by this amount. 

\par \citet{Char02} used the \textit{Hubble Space Telescope (HST)} to obtain the first detection of increased sodium absorption during transit of $0.023$ per cent at bandwidth 12\,{\AA} for the exoplanet HD\,209458b. \citet{Sing08} re-analysed these data and found  a signal of 0.11 per cent at bandwidth 4.4\,{\AA}. \citet{Nar05} made ground-based observations of the same planet, but found no significant sodium absorption. \citet{Snel08} (hereafter S08) improved on this analysis, finding absorption of 0.135 per cent at 1.5\,{\AA}; and \citet{Red08} detected sodium absorption of 0.067 per cent at bandwidth 12\,{\AA} in the atmosphere of HD\,189733b, also using ground-based spectroscopy. The sodium transit depths for HD\,209458b are around 3 times smaller than predicted by models with a cloudless atmosphere and solar sodium abundance \citep{Char02}, which shows that atomic sodium is depleted by some mechanism. Possible explanations include photoionization, condensation, or the presence of high clouds or haze in the atmosphere.   

\par Other atmospheric chemicals have been detected in exoplanet atmospheres. \citet{Tin07} used transmission spectroscopy with the \textit{Spitzer Space Telescope (Spitzer)} to detect the absorption signature of water in the atmosphere of HD\,189733b. \citet*{Swa08} used NICMOS on the \textit{HST} to detect methane and water in the atmosphere of the same planet. \textit{Spitzer} has also been used extensively for occultation spectroscopy -- the spectrum of the day-side of the planet can be obtained from the difference between the in-eclipse and out-of-eclipse spectra. While transmission spectroscopy probes the limb of the planet, occultation spectroscopy probes the day-side of the planet, and when these observations are at infrared wavelengths, temperature information can be obtained.

\par WASP-17b \citep{And10}, a hot Jupiter in Scorpius, is the least dense planet currently known, and is in a 3.74-d orbit around the star WASP-17, which has a \textit{V} mag of 11.6. WASP-17b is 0.05\,au from its F6-type host star, so is a pM class exoplanet \citep{For08} whose atmosphere is expected to have a thermal inversion layer. Anderson et al. (in preparation) finds, from an analysis of new data from \textit{Spitzer}, that the new parameters of the planet are 2.00\,$R_{\rm{J}}$ and 0.49\,$M_{\rm{J}}$. The planet has very low gravity, and the scale height of its atmosphere is approximately 5 times larger than HD\,209458b, so the sodium transit signal is expected to be correspondingly larger. However, if the atmosphere of WASP-17b is not sodium-depleted like HD\,209458b, the signal could be much larger. Similar observations are needed for many more hot Jupiters to improve models of their atmospheres.

\par We present here the first ground-based transmission spectroscopy of WASP-17b using the GIRAFFE spectrograph on the Very Large Telescope (VLT) in Chile. Our aim is to measure sodium absorption in and out-of-transit, and to detect any additional sodium absorption by the atmosphere of WASP-17b during the transit. We used the integral field unit-mode (IFU-mode) of FLAMES, which has the highest throughput of all the available instruments, allowing maximum collection of photons during the observations. The IFU-mode also allows simultaneous observation of a comparison star and sky background, without having to move the telescope.  

\par Details of the observations are given in Section~\ref{sec:obs}, analysis is described in Section~\ref{sec:ana}, the discussion is given in Section~\ref{sec:dis}, and Section~\ref{sec:con} concludes.

\section{Observations}\label{sec:obs}
 
\par We obtained 24 spectra of the WASP-17 system, eight during transit, using the 8-m VLT in Paranal, Chile. We used the GIRAFFE, fibre-fed spectrograph and the IFU-mode of FLAMES, mounted on Kueyen (UT2), to put IFUs on the the target star, a comparison star, and three sky regions. Each IFU is composed of a bundle of 21 fibres; 22 if the IFU includes a simultaneous calibration (simcal) fibre. Observations began at 22:56\,\textsc{ut} on 2009 June 20, and ended at 07:59\,\textsc{ut} on 2009 June 21. Mid-transit time of WASP-17b on 2009 June 21 was 04:38\,\textsc{ut}, and the transit duration was 4.4\,hours \citep{And10}. Start times of each exposure, which observing plate was used, mean airmass for that exposure and orbital phase are given in Table~\ref{tab:ot}.

\begin{table}
\small
 \caption{Exposure number, start times of target observations (in \textsc{ut}), observing plate used, mean airmass during exposure, and orbital phase (IT -- in-transit, OOT -- out-of-transit, Ing -- ingress, and Eg -- egress).}
 \label{tab:ot}
  \begin{tabular}{rrrrl}
  \hline
  No. & Time & Plate & Airmass & Phase \\ 
  \hline 
 1 & 22:56 & 2 & 1.5 & OOT \\
 2 & 23:17 & 2 & 1.4 & OOT \\
 3 & 23:38 & 2 & 1.3 & OOT\\
 4 & 23:58 & 2 & 1.2 & OOT \\ 
 5 & 00:19 & 2 & 1.2 & OOT \\
 6 & 00:46 & 1 & 1.1 & OOT\\
 7 & 01:07 & 1 & 1.1 & OOT \\
 8 & 01:28 & 1 & 1.0 & OOT\\
 9 & 01:48 & 1 & 1.0 & OOT\\ 
 10 & 02:09 & 1 & 1.0 & Ing\\
 11 & 02:52 & 1 & 1.0 & Ing\\
 12 & 03:13 & 1 & 1.0 & IT\\
 13 & 03:41 & 2 & 1.0 & IT\\
 14 & 04:02 & 2 & 1.1 & IT\\
 15 & 04:23 & 2 & 1.1 & IT\\
 16 & 04:43 & 2 & 1.1 & IT\\
 17 & 05:04 & 2 & 1.2 & IT\\
 18 & 05:25 & 2 & 1.3 & IT\\
 19 & 05:46 & 2 & 1.4 & Eg\\
 20 & 06:16 & 1 & 1.5 & Eg\\
 21 & 06:37 & 1 & 1.7 & OOT\\
 22 & 06:57 & 1 & 1.9 & OOT\\
 23 & 07:18 & 1 & 2.2 & OOT\\
 24 & 07:39 & 1 & 2.6 & OOT\\
 \hline
 \end{tabular}
\end{table}

Observations used the HR12 grating, which covers a wavelength range of $5821-6146$\,{\AA}. We obtained a resolution of 0.48\,{\AA} and a dispersion of 0.08\,{\AA} $\mathrm{pixel}^{-1}$. Target frames were exposed for 1000\,s, and simcal frames of 120\,s were taken separately in between each science frame. We could not observe the target while it passed through the zenith, so there is a gap between the 02:09 and 02:52 observations; this occurred during ingress, so does not affect our analysis. We obtained a total of 24 target frames; eight during transit, 12 out-of-transit, two during ingress, and two during egress. These frames contained spectra of WASP-17, the comparison star, and three separate sky regions. We also obtained 27 arc frames using the dedicated simcal fibre on the WASP-17 IFU, which were taken separately to avoid contamination of the target frames by leakage of light.

\par Seeing was very good, varying between 0.39 and 0.98 arcsec during the night, mainly remaining between 0.6 and 0.75 arcsec. Airmass was low for most of the observations -- see Table~\ref{tab:ot}. IFUs were placed on WASP-17 (an F6-type star with a \textit{V} mag of 11.6), and a comparison star, TYC2-6787-1398-1, an F7-type star with a \textit{V} mag of 11.3. The two stars had an angular separation of 13.6 arcmin. We also placed three IFUs on separate sky regions -- a total of five IFUs. There are two observation plates containing 15 available IFUs. Both plates were configured with the same five IFUs in identical positions, and plates were exchanged during the night to account for differential refraction across the field of view. 
Calibration was obtained from 3 daytime flats and a daytime arc for each observing plate, in addition to the simcal frames.  

\subsection{Data Reduction}\label{sec:dr} 
\par
\begin{figure}
\small
 \centering
 \includegraphics[width=7.5cm]{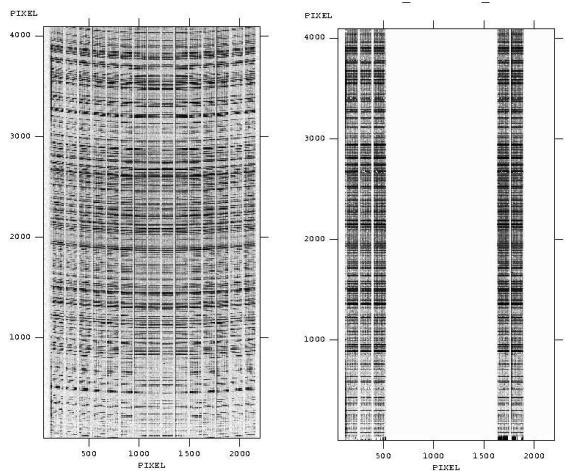}
 \caption{Left: the original arc image for plate 1. Right: the new arc image, after applying the straightening algorithm. There are some artefacts at the bottom of each column where no data exists. (The algorithm was developed for only the five IFUs we were using.)}
 \label{fig:bend}
\end{figure}

 The image of the slit on the CCD was curved in the vertical direction, and the position of each column on the slit varied in the horizontal direction -- see the left part of Fig.~\ref{fig:bend}, which shows a GIRAFFE thorium-argon arc image taken on the day of the observations. Each IFU on the image consists of about 130 columns of pixels, with wavelength read from bottom to top. The curvature in the vertical direction meant that some pixels which were part of a fibre in the middle of a column in the image were not part of the fibre at the top and bottom of the column. This was easily corrected for during extraction -- see subsection~\ref{sec:ext}.

\par The bend in the horizontal direction meant that adjacent pixels in a horizontal row of the image did not have the same wavelengths. We developed custom routines in Interactive Development Language (\textsc{idl})\footnote{\texttt{http://www.ittvis.com/ProductServices/IDL.aspx}}\label{fn:idl} to correct for this. We used the daytime arc images to measure the wavelength position for a set of 17 evenly spaced arc lines in each column of each of the five fibres we used. One column was used as a reference, then the other columns were interpolated on to the same wavelength scale, so that pixels of the same wavelength formed a horizontal row on the new, straightened image. Fig.~\ref{fig:bend} shows the original (left) and shifted (right) arc images for plate 1. We carried out this straightening step for each plate, using the corresponding daytime arc image. 

\subsection{Extraction}\label{sec:ext}
All frames were first bias-subtracted, using the overscan strip on each image. A master flat was created for each observing plate, using the median value of the three daytime flats. Then the corresponding straightening algorithm was applied to all the images obtained using that plate. Spectra were then extracted automatically using optimal extraction, which maximises the signal-to-noise ratio, and allows extraction of curved images \citep{Mar89}. The extraction routine also removed cosmic rays, although they had to be removed by hand in the sky frames, due to the low flux in those frames. The flat and arc spectra were extracted using the same profile as used for the data.  

\par The flat spectra were normalised, then each WASP-17 and comparison star spectrum was divided by its normalised flat. Arc calibration was performed using the simcal spectra, to track the drift in wavelength scale during the night of $\leq0.15$ pixels \citep{Pas02}. The wavelength scale was found from an eighth-order polynomial fit to 53 arc line positions; the standard deviation of the fit was $\leq0.005$\,{\AA}. The scale for an individual spectrum was found by interpolation between the two simcal spectra nearest in time to the exposure, using only spectra taken with the same plate. 

\par Before performing sky subtraction, we measured the radial velocity shifts between exposures for each IFU, to check fibre stability. All shifts were within the stated stability of 0.15 pixels in a 24-h period. The WASP-17 IFU and the IFU used for WASP-17 sky subtraction were separated by 1.8 arcmin; the comparison star IFU and the IFU used for comparison star sky subtraction were separated by 3.6 arcmin. Exposure 10 had a bright sodium emission line in both the sky and target frames which could not be subtracted without adding noise. This frame was discarded, but as it was taken during ingress, it was not needed for the analysis. Each image was then normalised and straightened by a straight line fit to a short length of the continuum  at each end of the spectrum; $5820.26-5844.45$\,{\AA} at the left-hand end, and $6028.33-6072.96$\,{\AA} at the right-hand end.

\par \sloppy Telluric correction was performed using a six-layer model of the Earth's atmosphere \citep{Nic88} and the {HITRAN} molecular database \citep{Rot05}, as implemented by \textsc{uclsyn}\footnote{\texttt{http://www.astro.keele.ac.uk/$\sim$bs/publs/uclsyn.pdf}}. The synthetic spectra were broadened to match the resolution of the observations, and a heliocentric correction was applied to the target images. Concentrating on the wavelength range $5880-5910$\,{\AA} (around the sodium lines), telluric $\mathrm{H}_{2}$O lines in each stellar spectrum were matched with those in a synthetic spectrum of similar airmass. The fit was judged by eye, then the stellar spectra were divided by the telluric spectra. This removed all visible traces of telluric contamination within the specified wavelength range. We estimate that any residual telluric $\mathrm{H}_{2}$O features remaining in this wavelength range are at a level of $\leq \pm 2.5$ per cent.

We also identified small telluric Na absorption lines in a high-resolution spectrum of WASP-17, identified on Fig.~\ref{fig:iscom} by two short arrows. These are of a similar strength to the weakest of the $\mathrm{H}_{2}$O lines \citep{Lund91}, a level in our WASP-17 spectra of 2.9 per cent. The resolution of our spectra was not high enough for us to identify the telluric Na lines, but, if they are present, they must have a similar strength in the both the WASP-17 and comparison star spectra. We did not attempt to subtract these telluric Na features, but the fact that the measured transit depths for the comparison star are consistent with zero at all bandwidths (see Fig.~\ref{fig:td}) shows that their variability is small. Statistical errors on every data point calculated from photon statistics were propagated through each stage of the data reduction. 

\section{Analysis}\label{sec:ana}
\par
\begin{figure}
 \centering
 \includegraphics[width=8cm]{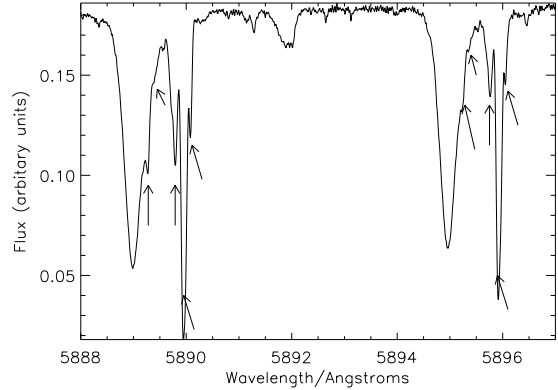}
 \caption{High-resolution, co-added spectrum of the WASP-17 system, taken with the HARPS spectrograph. Interstellar components of the sodium lines are marked with long arrows. The two short arrows indicate telluric Na absorption features. The stellar sodium lines are at 5889 and 5895\,{\AA}.}
 \label{fig:iscom}
\end{figure}

\begin{figure}
 \centering
 \includegraphics[width=8cm]{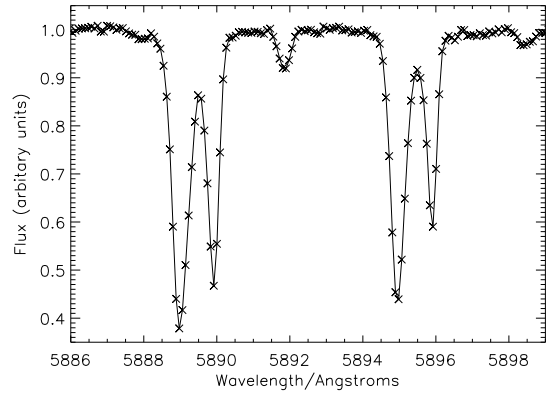}
 \caption{Normalised spectrum of WASP-17 taken at 22:56 \textsc{ut} on 2009 June 20. Sodium absorption lines produced by the star are at wavelengths 5889.01 and 5894.96\,{\AA}. Two broad interstellar Na features can be seen at wavelengths 5889.94 and 5895.96\,{\AA}.}
 \label{fig:wspec}
\end{figure}

 The position of each Na line component in each of the 24 exposures was measured by Gaussian fitting, and, when measuring flux, the mean position was used for the central line position, which was 5889.01\,{\AA} for the D$_{2}$ component, and 5894.96\,{\AA} for for the D$_{1}$ component. Fig.~\ref{fig:wspec} shows part of one of our extracted spectra, showing the area around the Na D lines. There are two broad interstellar Na features present in the spectra, which had a radial velocity shift of about $+47\,\mathrm{km\,s}^{-1}$ with respect to the stellar Na lines. A higher resolution, co-added HARPS spectrum of WASP-17 \citep{And10} shows that each interstellar feature is composed of at least four different components at different radial velocity shifts, one of which is unresolved and very close to the stellar line -- see Fig.~\ref{fig:iscom}. (Also indicated on the figure with two short arrows are telluric Na absorption lines.)

\par
\begin{figure}
 \centering
 \includegraphics[width=8cm]{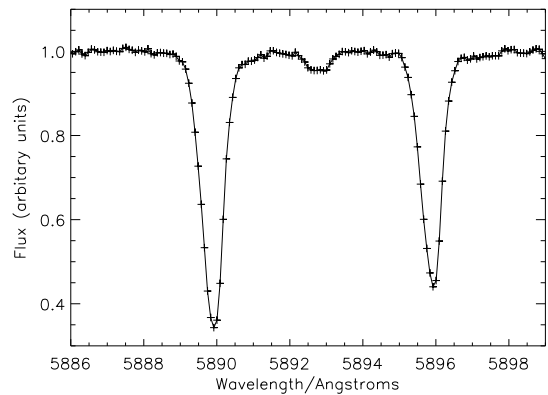}
 \caption{Normalised spectrum of comparison star TYC2-6787-1398-1 observed simultaneously with WASP-17. There are no interstellar features evident.}
 \label{fig:css}
\end{figure}

 The interstellar features are temporally stable, so are not originating from the WASP-17 system. They are likely to be caused by molecular clouds along the line of sight \citep{Wel10}. The comparison star does not show any Na interstellar features (see Fig.~\ref{fig:css}), although, at a distance of 280\,pc, it is closer to Earth than WASP-17 which is at at 380\,pc, so the clouds may be located between the two objects. The interstellar features do not vary during the transit; their presence will add a constant to both the in-transit and out-of-transit data, so should not alter the measured transit depths.

\par We found an $\sim 0.2$ pixel offset in the wavelength calibration between each plate -- the wavelength of absorption lines was $\sim 0.2$ pixels shorter in plate 1 images than in plate 2 images. This was removed by a custom \textsc{idl} routine. Our \textsc{idl} routine first shifted the wavelength of plate 2 exposures by the positive mean difference in wavelength between plates, then added a small shift to each of the four sets of exposures in order to line them up, and finally interpolated the data on to this new wavelength scale.

\par
\begin{figure}
 \centering
 \includegraphics[width=7cm]{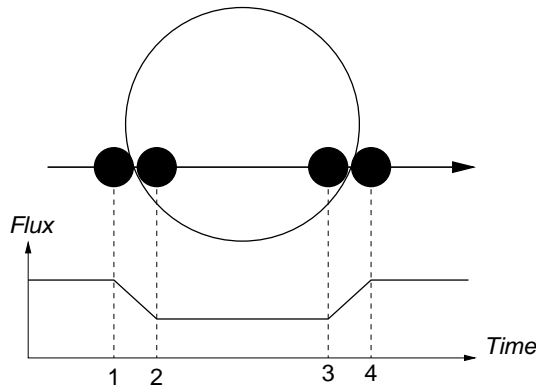}
 \caption{Transit diagram and related light curve. Ingress (between points 1 and 2) and egress (between points 3 and 4) frames were not used in our analysis.}
 \label{fig:tdg}
\end{figure}

 The flux in the Na lines was measured using bandwidths 0.75, 1.5, 3.0, 4.0, 5.0, and 6.0\,{\AA}; the ones at 1.5, 3.0, and 6.0\,{\AA} coinciding with the analysis of S08 to enable comparison. Half-bandwidths were centred on each line component; flux from each component was then added together. Ingress and egress frames were not used in the analysis -- only exposures taken between contact points 2 and 3 were taken to be in-transit, and only exposures taken before contact point 1 and after contact point 4 were taken to be out-of transit (see Fig.~\ref{fig:tdg}). Transit depth was calculated using: \[t_{D} = \frac{F_{\rm{in}}-F_{\rm{out}}}{F_{\rm{out}}}\] where $F$ is flux. $F_{\rm{in}}$ was calculated by taking the mean of the sum of the flux in each in-transit exposure for that bandwidth, and $F_{\rm{out}}$ was calculated in a similar way. 

\begin{figure}
 \centering
 \includegraphics[width=8cm]{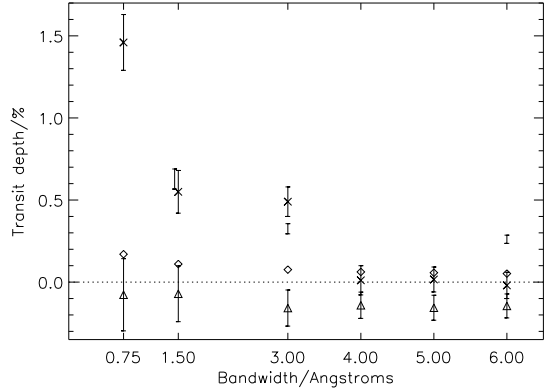}
 \caption{Transit depths for WASP-17b shown as crosses with error bars. S08 values for HD 209458b, scaled up by factors $4.2-5.1$, are shown as plain error bars; transit depths for the comparison star are shown as triangles with error bars. Diamonds represent the uncertainties due to photon noise.}
 \label{fig:td}
\end{figure}

Although we were able to measure significant transit depths using data from both observing plates, we found that discarding data from plate 1 led to lower uncertainties. We therefore give measured transit depths using only plate 2 data; out-of-transit data are exposures 1-5, and in-transit data are exposures 13-19 (see Table~\ref{tab:ot}). WASP-17b transit depth measurements are given in Table~\ref{tab:td} and Fig.~\ref{fig:td}. Transit depths are plotted as crosses with error bars, and diamonds show levels of photon noise. Predicted transit depths (S08 values for HD 209458b, scaled up by factors 4.2--5.1), are shown as plain error bars (the bar at 1.5\,{\AA} is slightly offset for clarity). We also measured transit depths in exactly the same way for the comparison star. These data are shown as triangles with error bars. We find no significant transit depth for the comparison star at any bandwidth, which shows that our systematic errors are small.

\captionsetup[table]{justification=raggedright}
\begin{table}
\small
\caption{Measured transit depths, per cent.}
 \label{tab:td}
  \begin{tabular}{rrr}
  \hline
  Band, {\AA} & WASP-17b & Comp. star \\ 
  \hline
  0.75  & $1.46\pm0.17$ & $-0.077\pm0.220$\\
  1.50 & $0.55\pm0.13$ & $-0.071\pm0.174$\\
  3.00 & $0.49\pm0.09$ & $-0.158\pm0.112$\\
  4.00 & $0.011\pm0.089$ & $-0.141\pm0.080$\\
  5.00 & $0.016\pm0.076$ & $-0.156\pm0.076$\\
  6.00 & $-0.019\pm0.076$ & $-0.145\pm0.073$\\ 
  \hline
 \end{tabular}
\end{table}

\begin{figure}
 \centering
 \includegraphics[width=8cm]{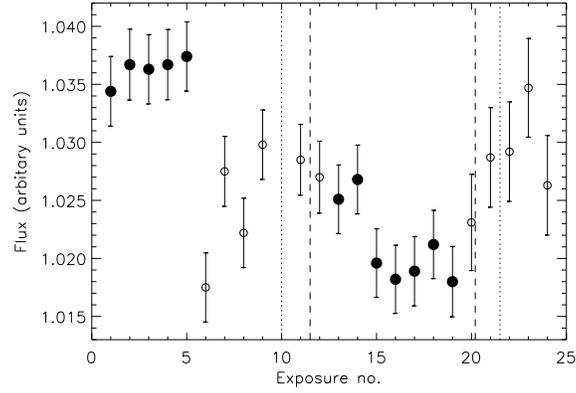}  
 \caption{Light curve of WASP-17b measured at 0.75\,{\AA}. Plate 1 exposures plotted as open circles; plate 2 data plotted as filled circles.}
 \label{fig:lc}
\end{figure}

We also show the 0.75\,{\AA} signal as a function of exposure number in Fig.~\ref{fig:lc}. The flux at half the bandwidth (0.375\,{\AA}) was measured at the centre of each line component; flux from each component was then added together. Data using observing plate 1 are plotted as open circles; data using observing plate 2 are plotted as filled circles. Exposures in between the dashed lines begin between contact points 2 and 3 (see Fig.~\ref{fig:tdg}), and exposures in between the dotted and dashed lines begin during ingress or egress. Exposure 10 is missing as we were unable to accurately measure its flux (see Sec.~\ref{sec:ext}).

\section{Discussion}\label{sec:dis}

We calculated predicted transit depths for WASP-17b, (shown on Fig.~\ref{fig:td} as plain error bars), by scaling up the S08 measured transit depths for HD\,209458b by the difference in scale height between the atmospheres of the two planets. This was calculated using the new planet mass and radius measurements and uncertainties (see Sec.~\ref{sec:in}), and is shown as a range. The scale height of a planet's atmosphere is given by: \[H = \frac{kT}{\mu g}\] $\textit{H}$ is about $500\,\mathrm{km}$ for HD\,209458b. The new mass and radius measurements give a scale height of $2090-2570\,\mathrm{km}$, $4.2-5.1$ times larger than that of HD\,209458b. The measured transit depth for WASP-17b of $0.55\pm0.13$ per cent at 1.5\,{\AA} is close to that predicted by scaling up the S08 measurements, and the transit depth of $0.49\pm0.09$ per cent at 3.0\,{\AA} is slightly higher than predicted. We find no significant transit depth at 6.0\,{\AA}, in contradiction to predictions.

\citet{Bro01} (hereafter B01) produced a model of the planetary absorption for HD\,209458b, for an ionized and non-ionized atmosphere -- see his fig.~18. We normalised these model spectra to exactly the same lengths of continuum as for our WASP-17b data, and measured the predicted transit depths for bandwidths 0.75, 1.5, 3.0, and 6.0\,{\AA}. These are given in Table~\ref{tab:predh}, together with the S08 measurements. 

\begin{table}
\small
\captionsetup[table]{justification=raggedright}
\caption{Measured and predicted transit depths (per cent) for HD\,209458b.}
 \label{tab:predh}
  \begin{tabular}{lrrrr}
  \hline
  ~ &   Band, {\AA}   &  ~  &  ~   & ~ \\ 
  ~ & 0.75  & 1.5 & 3.0 & 6.0 \\ 
\hline
S08 measurements        & -     & 0.135 & 0.070  & 0.056 \\
B01 Non-ionized model   & 0.262 & 0.225 & 0.193 & 0.165 \\
B01 Ionized model       & 0.166 & 0.144 & 0.126 & 0.111 \\
\hline
 \end{tabular}
\end{table} 

\begin{figure}
 \centering
 \includegraphics[width=8cm]{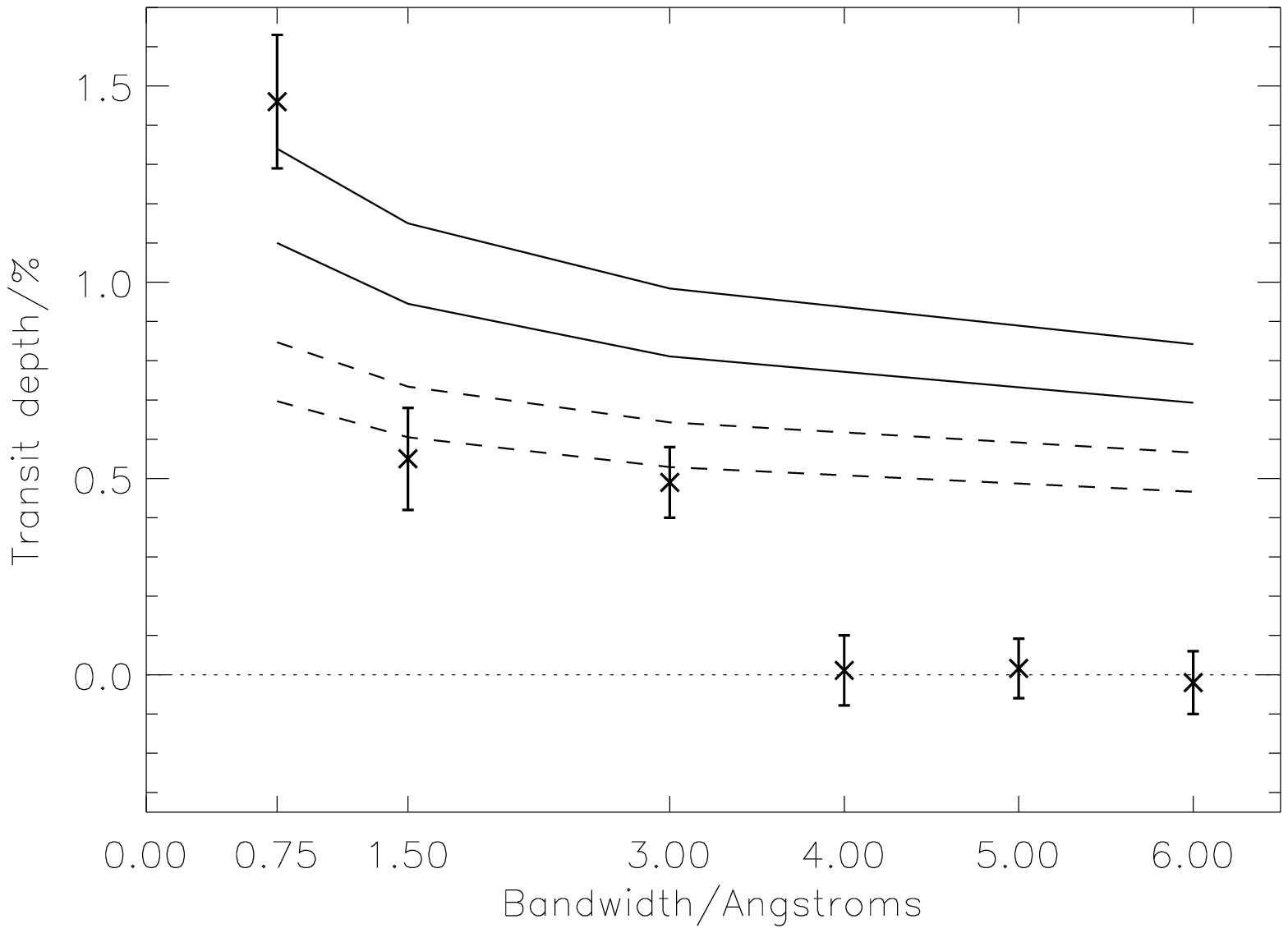}
 \caption{Bands show upper and lower bounds for transit depths predicted by B01 for WASP-17b -- the upper, solid band is for a non-ionized model atmosphere, and the lower, dashed band is for an ionized model atmosphere.}
 \label{fig:btdh}
\end{figure}

Mean sodium transit depths for WASP-17b predicted by scaling up the B01 models by the difference in atmospheric scale height are given in Table~\ref{tab:predw}. They are also plotted as bands, giving upper and lower bounds, in Fig.~\ref{fig:btdh}; the upper, solid band is for a non-ionized atmosphere model, and the lower, dashed band is for an ionized atmosphere model. Fig.~\ref{fig:btdh} also shows our transit depth measurements, plotted as crosses with error bars.

\begin{table}
\small
 \caption{Measured transit depths (per cent) for WASP-17b, and mean transit depths (per cent) predicted by the B01 model. (1) refers to a non-ionized model atmosphere; (2) refers to an ionized model atmosphere.}
 \label{tab:predw}
  \begin{tabular}{lrrrrrr}
  \hline
  ~ &   ~         &  Band, {\AA}   & ~        & ~        \\ 
  ~ & 0.75 & 1.5 & 3.0 & 4.0 & 5.0 & 6.0 \\ 
\hline
WASP-17b        & 1.46  & 0.55  & 0.49  & 0.01 & 0.02 & -0.02\\
B01 (1) & 1.22  & 1.05  & 0.90  & - & - & 0.77 \\
B01 (2)     & 0.77  & 0.67  & 0.59  & - & - & 0.52 \\
\hline
 \end{tabular}
\end{table} 

The transit depth for an ionized atmosphere predicted by the B01 data for HD\,209458b is 0.144 per cent at bandwidth 1.5\,{\AA}, slightly higher than that measured by S08. However, the predicted transit depths for the two wider bandwidths are almost twice as high as those measured by S08.

\par Our measurements of WASP-17b show a large transit depth at narrow bandwidths, a shallow transit depth at medium bandwidths, and no significant transit depth at bandwidths 4.0\,{\AA} or wider. Photo-ionization of sodium is expected to reduce the Na transit signal (see B01); \citet{Bar07} also modelled the wavelength-dependent transit radius of HD\,209458b and found that including photo-ionization in the model reduces the transit radius by a third at the wavelength of sodium. However, photo-ionization broadens the sodium absorption profile, which our measurements do not support.   

\par Depletion of heavy elements leads to a narrower sodium absorption profile, as the narrow component of the Na absorption line is formed at low pressures, high in the atmosphere, and the broad component is formed at high pressures, lower in the atmosphere. Depletion could be caused by the condensation of metals, including sodium, on the cold night-side of the planet. \citet*{Iro05} hypothesised that half of the limb is depleted in sodium by a factor of 3, due to condensation on the night-side of the planet, while \citet{For05} showed that the slant optical depth through the atmosphere of a hot Jupiter is up to 90 times greater than normal optical depth, and species with a small normal optical depth may mask Na absorption features. 

\par It is also possible that the broad Na absorption, which originates low in the atmosphere, is being masked by high cloud. \citet{For03} suggested that silicate and iron clouds high in the atmosphere could be an opacity source for sodium, and \citet{Pon08} concluded that their measurement of no significant absorption by Na or K at 30\,{\AA} bandwidths in the atmosphere of HD189733b was probably due to a haze of sub-micrometre sized particles high in the atmosphere, which masked the absorption signal. The details of the models used to calculate the effect of dust on alkali absorption line profiles have a large influence on the predicted spectra \citep{Joh08}, so it is likely that our observations have the potential to test directly models for the formation of dust at atmospheric temperatures and pressures typical for hot Jupiters and cool brown dwarfs \citep{Hell08}.

\par It seems likely that the Na depletion in WASP-17b is caused by a combination of the above factors; sodium is both ionized by the stellar flux, and depleted by condensation on the night-side. This would explain both the large transit depth at narrow bandwidths, and the lower than predicted transit depth at wider bandwidths. A layer of opaque cloud high in the atmosphere may explain the sharp cut-off in transit depth between 3 and 4\,{\AA}, or it could be caused by the masking of Na absorption features by other species high in the atmosphere.  

\par A transit depth measurement in such a narrow bandwidth is also affected by the rotation of the star. As the planet transits, it blocks the light from different regions of the rotating star. Light coming from parts of the star rotating towards the observer is blue-shifted, and light from parts of the star rotating away from the observer is red-shifted. This causes a spike in the radial velocity curve due to the orbital motion of the planet -- the Rossiter-McLaughlin effect. We are developing a model to account for this. 

\section{Conclusion}\label{sec:con}
On the night of 2009 June 20, we obtained spectra in and out-of-transit of  WASP-17b, a hot Jupiter in the Southern hemisphere with a large atmospheric scale height. We used the IFU-mode of FLAMES/GIRAFFE on the VLT to simultaneously obtain spectra of an F7-type comparison star and three sky background regions. We detected additional sodium absorption during the transit of $1.46\pm 0.17$ per cent at a measured bandwidth of 0.75\,{\AA} (9 $\sigma$), $0.55\pm 0.13$ per cent at 1.5\,{\AA} (4 $\sigma$), and $0.49\pm 0.09$ per cent at 3.0\,{\AA} (5 $\sigma$). We did not detect significant additional sodium absorption at 4.0, 5.0, or 6.0\,{\AA}. Our measurements show that WASP-17b, like HD\,209458b, is depleted in sodium compared to a model with a cloudless atmosphere and solar sodium abundance. The measured transit depths are best explained by a combination of Na ionization by stellar flux, Na condensation on the cold night-side, and either a layer of cloud or masking by opaque species high in the atmosphere. These measurements have provided useful information on the structure of the atmosphere of WASP-17b, which may apply to other hot Jupiters.   
\par The changes of observing plate during the night were a major source of systematic errors, and in future observations, only one plate should be used.  This is acceptable if no comparison star is observed, as there is no need to move the IFUs around to account for differential refraction over the field of view. We conclude that the IFU mode of FLAMES on the VLT is a very effective tool for obtaining transmission spectroscopy of extrasolar planets.

\subsection*{Acknowledgments}
We would like to thank David Sing for his comments on a poster based on these results that we found very useful in developing the Discussion and Conclusion sections of this paper. We would also like to thank the reviewer for their helpful comments. We thank Tom Marsh for the use of the programs \textsc{molly} and \textsc{pamela}\footnote{\texttt{http://deneb.astro.warwick.ac.uk/phsaap/software/}\label{fn:mar}}. PLW is supported by a Science and Technology Facilities Council postgraduate studentship.

%\end{multicols}
%\bibliography{mnrefs}
%\bibliographystyle{plainnat}

\end{document}